\def\be{\begin{equation}}
\def\ee{\end{equation}}
\def\bea{\begin{eqarray}}
\def\eea{\end{eqarray}}
\def\oneh{{\textstyle {1\over 2}}}

\documentclass{ws-p8-50x6-00}

\begin{document}

\title{(e,e$'$N) and (e,e$'$NN) experiments at NIKHEF, Mainz and JLab and
   open problems in one- and two-nucleon emissions}

\author{S.~Boffi}

\address{Dipartimento di Fisica Nucleare e Teorica,
Universit\`a degli Studi di Pavia,and \\ Istituto Nazionale di Fisica Nucleare, 
Sezione di Pavia, I-27100 Pavia, Italy}

\maketitle

\abstracts{A critical review is presented of recent data on direct one- and
two-nucleon emission obtained in electron scattering at NIKHEF, Mainz and JLab.
In the case of (e,e$'$p) reactions attention is focussed on extracting
spectroscopic factors, looking at medium effects on the bound nucleon form
factors, and investigating nuclear transparency. The possibility of obtaining
information on nucleon-nucleon correlations in (e,e$'$pp) and (e,e$'$pn)
reactions is also discussed.} 

\section{Introduction}
Electron scattering has been used for many years as a clean tool to explore
nuclear structure. In the one-photon-exchange approximation,
where the incident electron exchanges a photon of momentum $\vec q$ and energy
$\omega$ with the target, the response of atomic nuclei as a function of
$Q^2=\vert\vec q\vert^2-\omega^2$ and $\omega$ can nicely be separated because
the electromagnetic probe and its interaction are well under control. In
addition, in direct one- and two-nucleon emission one may access to the
single-particle properties of nuclei and nucleon-nucleon correlations,
respectively (see, e.g., Ref.~\cite{Oxford}).

In recent years, a great amount of data have been collected in laboratories in
Europe and the USA.

The Amsterdam Pulse Stretcher and Storage-ring facility (AmPS) at the Dutch
National Institute for Nuclear Physics and High-Energy Physics (NIKHEF) entered
the phase of regular operations in 1993 providing beams of longitudinally
polarized electrons of 900 MeV with stored currents of 150 mA and a
state-of-the-art experimental facility with high-density, polarized internal
targets. After a short period of intense and qualified work the AmPS facility
has been decommissioned in January 1999 due to budgetary constraints and
following the decision of the Dutch Research Council. Some part of the physics
programme, such as the two-nucleon emission programme, together with some parts
of the detectors, such as the Hadron3 detector originally designed to study
reactions with a small cross section, has been transferred to Mainz.

In Mainz electron scattering experiments are performed by the A1 collaboration
using a three-spectrometer setup to detect one or two charged particles in
coincidence with the scattered electron coming from the 855 MeV (polarized) beam
of the MAMI racetrack microtron. An upgrade to 1.6 GeV (MAMI-C) is under
development and the new complex is planned to run in Spring
2004.~\footnote{An overview of
approved experiments and accepted proposals can be obtained at the following web
site: 
http://wwwa1.kph.uni-mainz.de/A1/proposals.html.}
For the present purposes the experiments on
(e,e$'$pp) and (e,e$'$pn) proposed
under Refs.~\cite{A1-1-97,A1-5-98}, respectively, as well as the completed
experiment (e,e$'$p) under Ref.~\cite{Dieterich} will be considered.

In the USA, besides the MIT-Bates laboratory operating with its 1 GeV electron
beam, the Jefferson Laboratory (JLab) in Newport
News, Virginia, began conducting experiments in November 1995 with its
high-energy beams (up to 6 GeV at present, with possible upgrade to 12 GeV) of
the Continuous Electron Beam Accelerator Facility (CEBAF). One-proton emission
in electron scattering off a complex nucleus is studied in Halls
A and C. In Hall A cross sections of charged particles detected in coincidence
with the scattered electron can be measured with high precision with the
available two identical High-Resolution Spectrometers. In Hall C a variety of
experiments requiring high luminosity, but moderate resolution,
is possible with the High-Momentum Spectrometer and the Short-Orbit
Spectrometer.~\footnote{A full list of approved experiments  can be found at the following
web sites for Hall A and Hall C, respectively:
http://www.jlab.org/exp\_prog/generated/apphalla.html,
http://www.jlab.org/exp\_prog/generated/apphallc.html.}
Among others, of relevance here are the (e,e$'$p)
experiments~\cite{Gao,Liyanage,Malov,E-00-102} in Hall A, as well as the
proposal under Ref.~\cite{E-91-013} in Hall C and some corresponding
results~\cite{Abbott,Dutta}. The proposal by van den Brand {\sl et
al.\/}\cite{E-93-049} in Hall A will extend the Mainz data of
Ref.~\cite{Dieterich} from $Q^2 = 0.8$ to 4.0 GeV$^2$.

In this review attention
will be drawn to open problems more than to a successful interpretation of data.

\section{(e,e$'$N)}
In plane-wave impulse approximation (PWIA), i.e. neglecting
final-state interactions (FSI) of the ejected particle, the coincidence
(e,e$'$p) cross section in the one-photon exchange approximation is
factorized~\cite{Oxford,Report} as a product of the (off-shell) electron-nucleon
cross section $\sigma_{\rm eN}$ and the nuclear spectral density,
\be
S({\vec p}, E)=\sum_\alpha S_\alpha(E)\vert\phi_\alpha({\vec p})\vert^2.
\ee
At each removal energy $E$ the $\vec{p}$ dependence of $S({\vec p}, E)$ is given
by the momentum distribution of the quasi-hole states $\alpha$ produced in the
target nucleus at that energy and described by the (normalized) overlap
functions $\phi_\alpha$ between the target ($A$-particle) nucleus ground state and
the $(A-1)$-particle states of the residual nucleus. The spectroscopic factor
$S_\alpha$ gives the probability that the such a
quasi-hole state 
$\alpha$ be a pure hole state in the target nucleus. In an independent-particle
shell model (IPSM) $\phi_\alpha$ are just the single-particle states of the
model, and $S_\alpha=1 (0)$ for occupied (empty) states. In reality, the
strength of a quasi-hole state is fragmented over a set of single-particle
states due to correlations, and $0\le S_\alpha<1$.

With FSI such a factorization in
the cross section is no longer possible, but in the past
data were always organized in bins characterized by $\vert{\vec p}\vert$ and $E$
by on-line dividing the counting rates by (a model dependent) $\sigma_{\rm
eN}$. Thus a five-fold cross section was converted into a two-fold reduced
cross section and compared with a (distorted) spectral density  
$S^D(\vert{\vec p}\vert, E)$ 
including FSI described by an optical model potential. In the discrete spectrum of
the residual nucleus at each excitation energy
the $\vert{\vec p}\vert$ dependence of the reduced cross section is given by 
the corresponding model quasi-hole state and the normalization factor necessary
to adjust $S^D(\vert{\vec p}\vert, E)$ to data is interpreted  
 as the value of the spectroscopic factor extracted from experiment. 

\subsection{Spectroscopic factors and relativistic effects}
Two major findings came out of these studies. First, the valence quasi-hole
states $\phi_\alpha$ almost overlap the IPSM functions with only a
slight ($\sim10\%$) enlargement of their rms radius. Second, a systematic
suppression of the single-particle strength of valence states  as compared to
IPSM has been observed all over the periodic table. A quenching
of spectroscopic factors is naturally conceived in nuclear many-body theory in
terms of nucleon-nucleon correlations. However, model
calculations produce spectroscopic factors $S_\alpha$ much larger than those
extracted in low-energy (e,e$'$p) data. As an example, for
the p-shell holes in $^{16}$O a Green function approach to the
spectral density~\cite{Polls} gives  $S_{p_{1/2}} = 0.890$ and $S_{p_{3/2}} =
0.914$, while from experiment one has $S_{p_{1/2}} = 0.644$ and
$S_{p_{3/2}} = 0.537$. A recent reanalysis~\cite{Lapikas} of the
$^{12}$C(e,e$'$p) data at $Q^2\le 0.3$ GeV$^2$ has found a very substantial
reduction of the s- and p-shell strength by the factor $0.57\pm 0.02$.

In fact, the most general form of the coincidence cross section in the
one-photon-exchange approximation is the contraction of a lepton tensor
$L_{\mu\nu}$ with a hadron tensor $W^{\mu\nu}$ and involves nine structure
functions~\cite{Oxford,Report}, that describe the response of the target system to the
absorption of a longitudinal (L) or transverse (T) photon. 
Each individual structure function is a bilinear form of the
hadron current $J^\mu$, i.e.
\be
J^\mu = \int d\vec{r}\ \overline{\Psi}_{\rm f}(\vec{r})\, j^\mu(\vec{r})\,
e^{i\vec{q}\cdot\vec{r}}\,\Psi_{\rm i}(\vec{r}),
\label{eq:hadron}
\ee
where the charge-current operator $j^\mu(\vec{r})$ is responsible for the
transition from an initial state $\Psi_{\rm i}(\vec{r})$ (describing the motion
of the ejected nucleon in its initial bound state) to a final state $\Psi_{\rm
f}$ with the ejectile undergoing FSI with the residual nucleus. A complete
separation of the various structure functions with appropriate (out-of-plane)
kinematics would provide useful constraints when modelling the hadron current. 

In the nonrelativistic PWIA approach,
$\Psi_{\rm i}(\vec{r})$ is identified with 
$[S_\alpha]^{1/2}\phi_\alpha({\vec r})$ and
$\Psi_{\rm f}$ becomes a plane wave. When including FSI according to an approach
based on the distorted-wave impulse approximation (DWIA) and
followed in the past in connection with low-$Q^2$
data, $\Psi_{\rm f}$ is taken as a solution to a Schr\"odinger equation for the
ejectile in the field of the optical model potential produced by the residual
nucleus. Uncertainties arise because phase-shift equivalent potentials for
elastic proton-nucleus scattering do not in general give the
same result for $\Psi_{\rm f}$ in the nucleus interior. 

Ambiguities also arise in the definition of $j^\mu(\vec{r})$ that are related to
current conservation and off-shell behaviour~\cite{deForest,Kelly}. Under
quasi-free conditions the Coulomb gauge, where $J^3=(\omega/\vert{\vec
q}\vert)\,J^0$, and the Weyl gauge, where $J^0=(\vert{\vec q}\vert/\omega)
\,J^3$, give almost overlapping results. On the contrary, different expressions
of $j^\mu(\vec{r})$ derived making use of the Gordon
decomposition~\cite{deForest}, the socalled CC1 and CC2 currents, may give up to
10\% difference in the extracted spectroscopic factors~\cite{Udias}.

With increasing $Q^2$ and $\omega$ a relativistic approach becomes necessary.
The bound state wave function $\Psi_{\rm i}$ is now a four-spinor. In today's
calculations it is obtained within a Dirac-Hartree mean-field approximation of
the many-body problem.
Scalar ($S$) and vector ($V$) optical potentials in a Dirac equation for
$\Psi_{\rm f}$ describe the interaction of the ejectile with the rest of the
nucleus and are responsible for spinor distortion of the final state with
respect to the free case. This relativistic DWIA is thus based on 
(relativistic) IPSM wave functions. 

The presence of negative-energy components in the bound-state wave
function destroys factorization already in the relativistic PWIA cross
section~\cite{Gardner,Caballero}. However, the most important relativistic
effects are due to the replacement $E+M\to \tilde{E} + \tilde{M}= (E-V) +
(M+S)$, with $V$ positive and $S$ negative. This produces an overall reduction
due to the spinor normalization factor $(E+M)/2E$ and an enhancement of the low
components of the bound state wave function. This has
two consequences. First, the strength of the peak of the momentum distribution
at low $p$-values ($\le 300$ MeV) is reduced and correspondingly the
extracted spectroscopic factors are increased (by $\sim$ 10 -- 15\%, as already
estimated through the socalled Darwin term~\cite{Cannata}). Second, the
shape at high $p$-values ($\ge 300$ MeV) is also modified. In addition,
as the low components also depend on a ${\vec\sigma}\cdot{\vec p}$ factor, a
different behaviour is expected for the spin-orbit partner shells,
$j=l\pm\oneh$, with a major sensitivity in $R_{LT}$ and the left-right
asymmetry, larger for the jack-knifed $j=l-\oneh$ (e.g. $p_{1/2}$ in $^{16}$O)
than for the stretched $j=l+\oneh$ (e.g. $p_{3/2}$) (see
Ref.~\cite{Caballero98}). 

\begin{figure}[t]
\vspace*{-0.3cm}
\hspace*{2cm}\epsfig{file=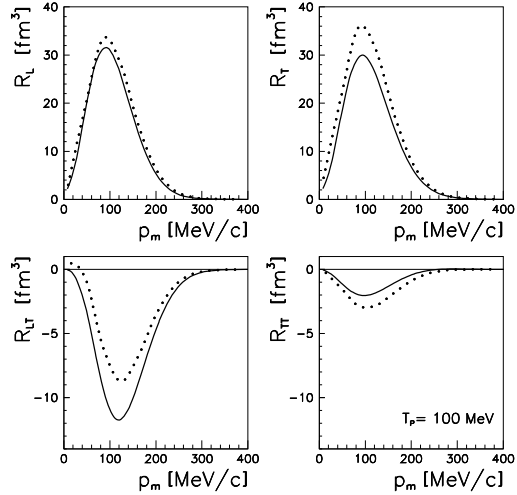,width=18pc}
\vspace{-0.5cm}
\caption{Separated structure functions for the reaction
$^{16}$O(e,e$'$p)$^{15}$N$_{\rm g.s.}$ with an emitted proton with 100 
MeV kinetic energy. Solid (dashed) lines are relativistic (nonrelativistic)
results (from Ref.~{\protect\cite{Meucci}}).}
\label{fig:Meucci}
\vspace{-0.5cm}
\end{figure}

The relevance of genuine relativistic effects has recently been
investigated~\cite{Meucci} in a consistent comparison between nonrelativistic
and relativistic calculations. Significant relativistic effects,
especially in $R_T$ and $R_{LT}$, are found already for a proton kinetic energy
as low as 100 MeV (Fig.~{\ref{fig:Meucci}). 

In the kinematics of Ref.~\cite{Gao}, i.e. $Q^2=0.8$ GeV$^2$ and $\omega=439$
MeV, relativity is quite important to describe the $^{16}$O
data~\cite{Udias99,Kelly99}. It 
is remarkable that the extracted spectroscopic factors are in this case much
larger than those obtained at lower $Q^2$, i.e. $S_{p_{1/2}} = 0.73$ and
$S_{p_{3/2}} = 0.71$ in Ref.~\cite{Udias99}, and $S_{p_{1/2}} = 0.72$ and
$S_{p_{3/2}} = 0.67$ in Ref.~\cite{Kelly99}. However, one has also to observe
that the data of Ref.~\cite{Gao} for the first time are given directly in terms
of cross sections, not (model dependent) reduced cross sections. 

\subsection{Transparency}
The (e,e$'$p) reaction under quasi-free kinematics is also a valuable
tool to study nucleon propagation in the nuclear medium. Data are
available for $Q^2$ up to 7 GeV$^2$ on a variety of 
target nuclei~\cite{propagation}. A systematic study has recently been
accomplished~\cite{Abbott} for proton kinetic energies in the range between 300
and 1800 MeV that includes the minimum of the nucleon-nucleon
total cross section and its rapid rise above the pion production threshold.
These features are partially reflected in the energy-dependent attenuation
of the proton flux. This attenuation is measured by the transparency ratio  
\be
T = \frac{\displaystyle \int_{\Delta E} {\rm d}E \int_{\Delta\vec{p}} {\rm
d}\vec{p}\, \sigma_{\rm exp}(\vec{p},E)}
{\displaystyle \int_{\Delta E} {\rm d}E
\int_{\Delta\vec{p}} {\rm d}\vec{p}\, \sigma_{{\rm PWIA}}(\vec{p},E)},
\label{eq:transparency}
\ee
where $\Delta E$ and $\Delta\vec{p}$ define the range of missing energy and
momentum explored by the experiment. In the data analysis a factorized
expression is assumed both for $\sigma_{\rm exp}$ and $\sigma_{\rm PWIA}$ and,
apart from kinematics, relativity is ignored in the calculation of $\sigma_{\rm
PWIA}$. Thus $T$ is a heavily model dependent quantity.
However, the role of genuine attenuation of FSI with increasing energy must be
understood before studying other mechanisms, such as e.g. color 
transparency.  

In fact, FSI at high energy are described in the eikonal or Glauber
approximation~\cite{Frankfurt}. The multiple scattering of the ejected proton
can also be described semiclassically within the intranuclear cascade
model~\cite{Golubeva}. This model has the advantage of directly implementing the
detector acceptances that limit the range of $\Delta E$ and $\Delta\vec{p}$. In
contrast, the eikonal approximation allows
for a detailed analysis of the contribution of each shell to the integrated
transparency as well as to its angular distribution~\cite{Golubeva}. It is
remarkable that quite similar results are obtained with the two approaches in
good agreement with experiment by simply assuming a full occupation of the
IPSM states. This is confirmed in the Glauber approach of Ref.~\cite{Frankfurt}
where no quenching of $S_\alpha$ is found and the role of short-range
correlations is shown to be insignificant.

A possible $Q^2$-dependence of spectroscopic factors, jumping from values around
0.6--0.7 at low $Q^2$ to 
unity at $Q^2\ge 1$ GeV$^2$, simply means that
something is not under control in either experiment or theory or both. One
certainly should measure exclusive cross sections instead of providing 
(model dependent)
reduced cross sections or transparencies. On the theory side a consistent
approach is desirable, where relativity and FSI interactions are appropriately
taken into account in calculating an unfactorized cross section. In particular,
spin-orbit effects should be investigated.


%
\begin{figure}[b]
\vspace*{-0.3cm}
\hspace*{2.5cm}\epsfig{file=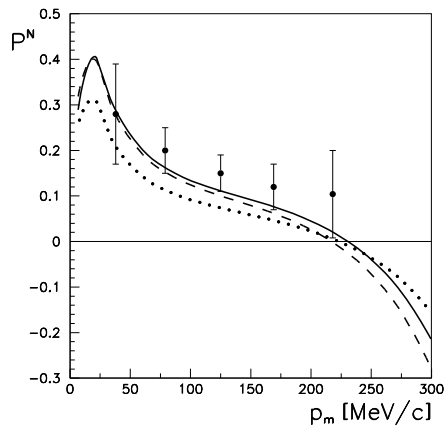,width=16pc}
\vspace{-0.5cm}
\caption{The induced polarization of the p$_{3/2}$ proton hole in the
$^{12}$C(e,e$'{\vec{\rm p}}$) reaction as a function of the missing momentum
$p_m$. Data from Ref.~{\protect\cite{Woo}}.
Solid (dotted) line with the EDAD1 (EDAI-C) optical potential of
Ref.~{\protect\cite{Cooper}}. Dashed line obtained with EDAD1 after removing the
negative-energy components in the bound state (from
Ref.~{\protect\cite{Meucci}}).} 
\label{fig:Meuccipol}
\vspace{-0.5cm}
\end{figure}

\subsection{Polarization observables}
The measurement of proton polarization is simpler than the separation of the
(polarized) structure functions and less affected by
experimental errors, as it is obtained through the determination of asymmetries.
With polarized incident electrons and polarized recoiling nucleons
with spin directed along $\hat{s}$ the cross section
for a proton detected in the solid angle ${\rm d}\Omega_{\rm p}$
in coincidence with the electron of energy $E'$ scattered in 
${\rm d}\Omega'$ can also be written 
\be
\frac{{\rm d}^3\sigma}{{\rm d}E'\,{\rm d}\Omega'\,{\rm d}\Omega_p} 
= \oneh\sigma_0\left[1+ \vec{P}\cdot\hat{s} 
+ h\left( A + \vec{P}\,'\cdot\hat{s}\right)\right],
\ee
where $\sigma_0$ is the unpolarized differential cross section, $\vec{P}$ the
outgoing nucleon polarization, $h$
the electron helicity, $A$ the electron analyzing power, and $\vec{P}\,'$ the
polarization transfer. In coplanar kinematics, only the component $P^N$ of $\vec
P$ normal to the scattering plane survives, and ${\vec P}\,'$ lies within the
scattering plane with components ${P'}^L$ and ${P'}^S$, parallel (longitudinal)
and perpendicular (sideways) to the outgoing proton momentum, respectively. 

Without FSI, $P^N=0$. Therefore $P^N$ is a good candidate to look at when
studying nuclear transparency, as its $Q^2$ dependence reflects the energy
dependence of FSI. A first experiment has been performed on $^{12}$C at
MIT-Bates~\cite{Woo} for $(\omega,q)=(194, 756)$
MeV. Relativistic DWIA results are indeed sensitive to the model used to
simulate FSI. In Fig.~{\ref{fig:Meuccipol} they are 
compared with data using two versions of an optical model potential based on 
empirical effective interactions~\cite{Cooper}. The role of relativity is
appreciated by removing the negative-energy components
in the bound state (dashed line), while sensitivity to FSI is shown by the
difference between solid and dotted lines. In particular, these data are
dominated by the real part of the spin-orbit optical
potential~\cite{Woo,Vignote}. 

\begin{figure}[b]
\vspace*{-0.3cm}
\hspace*{2cm}\epsfig{file=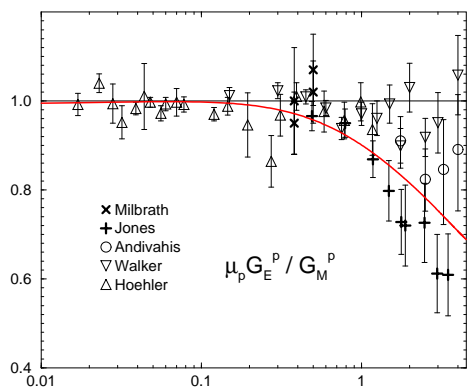,width=18pc}
\vspace{-0.5cm}
\caption{The ratio $G_E^p/G_M^p$ rescaled by the theoretical proton magnetic
moment $\mu_p$ in a (covariant) point-form approach with the
Goldstone-boson-exchange quark model, compared to data as indicated
(from Ref.~{\protect\cite{Wagenbrunn}}).} 
\label{fig:ge}
\vspace{-0.5cm}
\end{figure}

Polarization transfer is a powerful technique to investigate the behaviour of
the nucleon in vacuum and in the nucleus. The ratio 
${P'}^S/{P'}^L$ for a free proton is directly proportional to the ratio between
the electric and 
magnetic form factors through a factor depending on the electron kinematics
only: 
\be
\frac{G^p_E}{G^p_M} = - \frac{{P'}^S}{{P'}^L}
\frac{E+E'}{2M}\,\tan(\oneh\theta).
\label{eq:ratio}
\ee
The two recoil polarization components were measured simultaneously up
to 6 GeV$^2$ at JLab~\cite{Jones} with the surprising result that the ratio
(\ref{eq:ratio}) decreases linearly with $Q^2$ starting around
0.8 GeV$^2$. This implies that $G^p_E$ decreases much faster than the
dipole form factor and nonrelativistically this means
that the electric charge distribution in the proton extends to larger distances
than the magnetization one. This trend is well reproduced in a chiral
constituent-quark model where covariance is ensured in a point-form
approach~\cite{Wagenbrunn} (Fig.~\ref{fig:ge}).

\begin{figure}
\vspace*{-0.3cm}
\hspace*{2cm}\epsfig{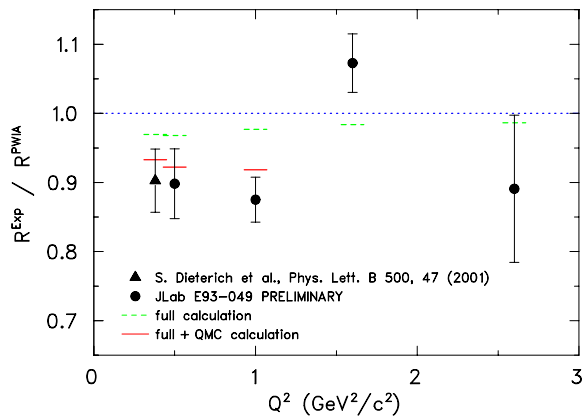}
\caption{The observed super-ratio of Eq.~(\ref{eq:superratio}) compared with the
calculated one with (solid) and without (dashed) medium modified form factors.
See text (courtesy of S.~Dieterich).} 
\label{fig:superratio}
\vspace{-0.5cm}
\end{figure}

For a nucleon embedded in the nuclear medium one would expect some
effects on the spatial distribution of its constituents due to the close
proximity of the other nucleons. To observe such effects the ratio $R_T/R_L$ has
been proposed, as for a free nucleon it is proportional
to $G_M^p/G^p_E$. However, it has proven to be a very difficult 
task to disentangle changes of the nucleon structure from other conventional
nuclear effects, such as meson-exchange currents, isobar configurations and FSI,
because the nucleon form factors of a bound nucleon are not observable (see
Ref.~\cite{Dutta} and references therein). 

Polarization observables are less sensitive to
systematic uncertainties and model ambiguities than
cross sections. In a first experiment~\cite{Malov} on $^{16}$O the ratio
${P'}^S/{P'}^L$ at $Q^2=0.8$ GeV$^2$ has been found in good agreement with
calculations based on the free proton form factors with an experimental
uncertainty of about 18\%. A more precise measurement involves the super-ratio
\be
R = \frac{\vert{{P'}^S}/{{P'}^L}\vert_A}{\vert{{P'}^S}/{{P'}^L}\vert_p}
\label{eq:superratio}
\ee	
obtained by dividing the ratio ${P'}^S/{P'}^L$ observed in proton knockout off a
nucleus $A$ by that observed in electron-proton elastic scattering. The
observed $R$ for a target $^4$He nucleus is compared in
Refs.~\cite{Dieterich,E-93-049} with results obtained with a variety of
different models, including medium-modified proton form factors. The PWIA
calculation (dotted horizontal line in 
Fig.~\ref{fig:superratio}) serves as baseline. The full relativistic
calculation~\cite{Vignote}
predicts a reduced ratio, but cannot fully account for the measurement.
The result up to $Q^2=1$ GeV$^2$ is in favour of
some density-dependent form factor modification as predicted by the quark-meson
coupling model of Ref.~\cite{Lu}. A proposal for a
similar analysis on $^{16}$O is under discussion~\cite{PR-01-013}.

\section{(e,e$'$NN)}
There is accumulating evidence for enhanced (e,e$'$p) transverse strength of
non-single particle origin at high missing energies~\cite{Ulmer,Dutta}. New
data~\cite{Liyanage} on $^{16}$O(e,e$'$p) cross section and separated responses
at $Q^2=0.8$ GeV$^2$ indicate a clear 1s peak at 40 MeV missing energy at small
missing momenta $p_m$, but at larger $p_m$ there is no peak and the DWIA
knockout cross section becomes much smaller than the data. For $p_m>200$ MeV the
cross section is almost constant, (e,e$'$pp) and (e,e$'$pn) contributing by
about one half of the measured cross section~\cite{Janssen}. Besides testing the
limits of the single-particle model in one-nucleon knockout~\cite{E-00-102},
experiments are also proposed to directly observe two nucleons ejected in
coincidence with the scattered electron.

Exclusive two-nucleon emission by an electromagnetic probe has been proposed
long time ago~\cite{Gottfried} to study nucleon-nucleon correlations. Due to 
the difficulty of measuring exceedingly small cross sections in
triple coincidence, only with the advent of high-duty-cycle electron beams has a
systematic investigation become possible. At
present, only a few pioneering measurements have been carried
out~\cite{Onderwater97,Onderwater98,Starink,A1-1-97,Rosner}, but
the prospects are very encouraging.

The general theoretical framework involves the two-hole spectral
density~\cite{Boffi,Oxford,GP}, whose strength gives the probability of removing
two nucleons from the target, leaving the residual nucleus at some excitation
energy. Integrating the two-hole spectral density over the energy of the
residual nucleus one obtains the two-body density matrix incorporating
nucleon-nucleon correlations. The triple coincidence cross section is again a
contraction between a lepton and a hadron tensor, which contains the
two-hole spectral density through bilinear products of hadron
currents $J^\mu$ of the type (\ref{eq:hadron}) suitably adapted to a final state
with two ejected nucleons. FSI in principle require the solution of a three-body
problem in the continuum. Thus one usually approximate FSI by an attenuated 
flux of each ejectile due to an optical model potential. 

Even without FSI the two-hole spectral density is not factorized in the
triple coincidence cross section. This makes a difficult task to extract
information on correlations from data, and models are required to investigate
suitable kinematic conditions where the cross section is particularly sensitive
to correlations. A priori one may envisage that two-nucleon knockout is due
to one- and two-body currents. Of course, one-body currents are only effective  
if correlations are present so that the nucleon interacting with the incident
electron can be knockout together with another (correlated) nucleon. In
contrast, two-body currents, typically due to meson exchanges and isobar
configurations, lead naturally to two-nucleon emission even in an
independent-particle shell model. 

Two-body currents are mainly transverse and preferentially involve a
proton-neutron pair. Thus reactions like ($\gamma$,pn) and (e,e$'$pn) are
particularly sensitive to their effects. In contrast, (e,e$'$pp) reactions,
where two-body currents play a minor role, are better suited to look for
correlations, and resolution of discrete final states has been shown to provide
an interesting tool to discriminate between contributions of different
mechanisms responsible for two-nucleon emission~\cite{Giusti98}.

The shape of the angular distribution of the two emitted nucleons mainly
reflects the momentum distribution of their c.m. total angular momentum $L$
inside the target nucleus. When removing two protons from the $^{16}$O ground
state, the relative $^1S_0$ wave of the two protons is combined with $L=0$ or 2
to give $0^+$ or $2^+$ states of the residual $^{14}$C nucleus, respectively,
while the relative $^3P$ waves occur always combined with a $L=1$ wave function
giving rise to $0^+, 1^+, 2^+$ states. Combining the reaction description of
Ref.~\cite{GP} with the many-body calculation of the two-particle spectral
function in $^{16}$O of Ref.~\cite{Geurts}, in Ref.~\cite{Giusti98} the cross
section for the $0^+$ ground state, and to a lesser extent also for the first
$2^+$ state of $^{14}$C, was shown to receive a major contribution from the
$^1S_0$ knockout. Such transitions are therefore most sensitive to short-range
correlations. This is indeed the case, as seen in two exploratory studies
performed at NIKHEF~\cite{Onderwater97,Onderwater98}, and confirmed in
Ref.~\cite{Starink}. As the calculations give significantly different results
for different correlation functions, precise data could give important
constraint when modelling
the off-shell behavior of the nucleon-nucleon potential.

Superparallel kinematics has been preferred at Mainz~\cite{Rosner}, with one
proton ejected along the virtual photon direction and the other in the opposite
direction. In this kinematics only the pure longitudinal and pure
transverse structure
functions occur in the cross section, and a Rosenbluth L/T separation becomes
possible. The effect of two-body currents is further suppressed by looking at
the longitudinal structure function that is most sensitive to short-range
correlations. The data are still preliminary and require further analysis before
a fully reliable comparison with calculations can be done. Nevertheless they
show distinctive features predicted by calculations~\cite{Giusti98,Ryckebusch}.

Tensor correlations are expected to play a major role in (e,e$'$pn) reactions
where, however, the proton-neutron pair is ejected by a much more
complicated mechanism involving two-body currents. In the superparallel
kinematics of the proposed Mainz experiment~\cite{A1-5-98} with
$(\omega,q)=(215,316)$ MeV the predicted cross sections for (e,e$'$pn) are about
one order of magnitude larger than the corresponding cross sections for
(e,e$'$pp) reactions~\cite{GP99}. This enhancement is partly due to
meson-exchange 
currents and partly to tensor correlations. Quite different results are
predicted depending on these correlations being included or not. An accurate
determination of the two-hole spectral density is thus most desirable in order
to disentangle the effects of two-body currents from those of nuclear
correlations. 

Experimentally, additional and precise information will come from measurement of
the recoil polarization of the ejected proton in either (e,e$'$pp) or
(e,e$'$pn). Resolving different final states is a precise filter to disentangle
and separately investigate the different processes due to correlations and/or
two-body currents. The general formalism is available~\cite{GP00} and has been
extended to study polarization observables also in the case of two nucleons
emitted by a real photon~\cite{GP01}.

\section{Conclusions}

The advent of high-energy continuous electron beams coupled to high-resolution 
spectrometers has opened a new era in the study of basic nuclear properties such
as single-particle behaviour and nucleon-nucleon correlations by means of one-
and two-nucleon emission. In parallel new
relativistic theoretical approaches have been developed showing that
relativistic effects are most important and affect the interpretation of data
even at moderate energies of the emitted particles. A striking feature coming
out of the present analysis of (e,e$'$p) world data is an apparent $Q^2$
dependence of the extracted spectroscopic factors. This is clearly not
acceptable and indicates that something is missing in the theoretical treatment
of the reaction mechanism. However, it is also rather important to measure cross
sections directly without introducing a model-dependent treatment of data as was
done in the past at low $Q^2$ in order to produce reduced cross sections or at
high $Q^2$ when looking at the nuclear transparency. 

Polarization observables are most useful to gain information about 
hadron currents and final-state interactions. As the normal recoil polarization
$P^N$ vanishes without FSI, its $Q^2$ dependence in exclusive (e,e$'$p)
reactions deserves a detailed study in connection with the problem of nuclear
transparency and, ultimately, of color transparency. Measuring the components of
the induced (transfer) polarization, on the other side, gives important
information about possible medium modification of the nucleon electromagnetic
form factors. Ultimately, a complete determination of the scattering amplitudes is
only possible with polarization measurements.

Exclusive experiments with two-nucleon emission in electron scattering require
triple coincidences with three spectrometers. This is now possible and the first
experiments have been performed. By an appropriate selection of 
kinematic conditions and  specific nuclear transitions, it has been
shown that data are sensitive to nuclear correlations. In turn these 
strictly depend on the nucleon-nucleon potential. Therefore, two-nucleon
emission is a promising field
deserving further investigation both experimentally and theoretically in
order to solve a longstanding problem in nuclear physics. 

In conclusion, it is clear that the field of electron scattering on complex
nuclei offers a wide spectrum of still open problems in understanding the
nuclear behaviour.

\section{Acknowledgements}
I would like to thank the organizers for a most pleasant workshop and their warm
hospitality. I also would like to thank
the Institute for Nuclear Theory at the University of Washington for its
hospitality and the Department of Energy for partial support during the
completion of this work.

\end{document}